\documentclass[aps,pre,twocolumn,groupedaddress,showpacs]{revtex4}
\usepackage{amsmath}
\usepackage{graphicx}
\usepackage{color}
\newcommand{\be}{\begin{equation}}
\newcommand{\ee}{\end{equation}}
\newcommand{\bea}{\begin{eqnarray}}
\newcommand{\eea}{\end{eqnarray}}

\newcommand{\dd}{\partial}

\begin{document}


\title{Forced transport of deformable containers through narrow constrictions}


\author{ Remy Kusters$^1$, Thijs van der Heijden $^{1}$, Badr Kaoui $^{1,2}$, Jens Harting$^{1,3}$ and Cornelis Storm $^{1}$\\
\small $^1$Department of Applied Physics, Eindhoven University of Technology, Den Dolech 2,\\ 
\small 5600MB Eindhoven, The Netherlands\\ \small $^2$ Theoretical Physics I, University of Bayreuth, D-95447 Bayreuth, Germany \\
\small $^3$ Faculty of Science and Technology, Mesa+ Institute, University of Twente, \\
\small 7500AE Enschede, The Netherlands \\}


\date{\today}

\begin{abstract}
We study, numerically and analytically, the forced transport of deformable containers through a narrow constriction. Our central aim is to quantify the competition between the constriction geometry and the active forcing, regulating whether and at which speed a container may {\em pass} through the constriction and under what conditions it gets {\em stuck}. We focus, in particular, on the interrelation between the force that propels the container and the radius of the channel, as these are the external variables that may be directly controlled in both artificial and physiological settings. We present Lattice-Boltzmann simulations that elucidate in detail the various phases of translocation, and present simplified analytical models that treat two limiting types of these membrane containers: deformational energy dominated by the {\em bending} or {\em stretching} contribution. In either case we find excellent agreement with the full simulations, and our results reveal that not only the radius but also the length of the constriction determines whether or not the container will pass.
\end{abstract}


\maketitle

\section{Introduction}

Membrane-enclosed vesicles are the principal carriers used in intracellular protein trafficking. Moreover, because of their intrinsic biocompatibility and flexibility they are becoming an increasingly common motif in drug delivery, for instance in transdermal applications, as well as in microfluidic production and processing \cite{Lopez2004, Lopez2005}. In each of these settings, vesicles frequently encounter narrow passages: geometric constrictions that force them to change shape dramatically in order to pass. While driving forces such as pressures (possibly osmotic), fluid flow, directly exerted forces from molecular motors or external fields may promote passage, the required changes in shape generically result in energetic barriers to translocation and the eventual (non-) passage is thus determined by a subtle balance of forces originating from various physical sources, as well as by the geometry of the constriction. 

Specific examples of the channel passage problem are encountered in microfluidic devices in medical diagnostics \cite{Shelby2003, Suresh2007, Fedosov2014} and the fabrication of microgel capsules \cite{Seiffert2010}. Experimental work on red blood cells \cite{Shelby2003} and polymeric capsules \cite{ Risso2006} has shown that changes in mechanical properties and cell radius determine the passage of the container and in some cases may induce capillary blockage \cite{Leong2011, Zhou2007}. Previous theoretical work has extensively studied the transient dynamics of elastic capsules in both cylindrical and rectangular constrictions \cite{Queguiner1997, Park2013, Kuriakose2011},  the production of smaller vesicles \cite{Bertrand2012} and the translocation of vesicles through narrow pores \cite{Linke2006}.  Experimental work on transfersomes has demonstrated that ultra-flexible artificial liposomes roughly 500 nm in diameter may pass through pores as small as 50 nm virtually unobstructed, lending clear credibility to the paramount importance of membrane bending energies in this process \cite{Cevc1995, Cevc1998}.

Our own interest in the problem is further sparked by the regulatory use of recycling endosomes in dendritic spines: large lipid bilayer vesicles are actively directed by myosin motors \cite{Desnos2007} into, and out of, a long thin neck that connects the functional domain of a dendritic spine to the dendritic shaft. These vesicles are thought to serve dual purposes: they actively transport membrane-bound glutamate receptors to the functional domain, but, when stuck inside the neck, may also serve as a physical barrier \cite{Kennedy2006, Derkach2007, Kusters2013} that helps retain proteins inside the spine's head compartment \cite{Park2006, Wang2008} - not unlike the manner in which a cork serves to keep wine inside the bottle.

While this prior work has laid important foundations for our understanding of the process of vesicle translocation, much is still unclear. In particular, the dynamics of the translocation process still poses some open questions: How fast is the container transported through the constriction? When does it cease to translocate, and are typical molecular force levels sufficient to effect translocation in physiological settings? We focus in particular on those physical variables that cells have some control over: motor activity and constriction geometry. We also consider the dependence on the membrane's mechanical moduli which in synthetic settings such as vesicle production or extrusion may be controlled and optimized. Our principal interest, however, lies with the basic competition between the constriction geometry and the active forcing: how the shape, length and radius of the constriction and the force regulate the transport of a deformable container through a narrow constriction. We model the energetics of the transportation of deformable containers through a narrow constriction in a Lattice-Boltzmann simulation, combined with an immersed boundary method and a finite element method, and a simplified theoretical model. Our theoretical model may be applied to a wide range of deformable containers, but here we restrict ourselves to the discussion of two limiting cases: highly {\em stretchable} containers and inextensible membrane containers, to which we will refer as {\em capsules} and {\em vesicles} respectively.  Our analysis reveals a generic phase behavior of the {\it Stuck} and {\it Pass} regimes as function of the applied force, relative size of the constriction and the mechanical properties, in both the vesicle and capsule limits. In addition to infinitely long constrictions, we model the effect of a finite constriction length and show that the deformation energy and thus the minimal force necessary to get the capsule through the constriction significantly decreases for decreasing neck length. 

This paper is organized as follows: Section II discusses the Lattice-Boltzmann simulation and presents the key results of our simulation. In Section III we introduce our simplified theoretical model and outline calculations of the two limiting cases (stretch vs bend dominated containers). In Section IV we compare and discuss the results of our theoretical model with the simulation results and present our main conclusions.

\section{Lattice-Boltzmann simulations}

In this section we outline our three-dimensional Lattice-Boltzmann simulations for the deformable container and present our main results. In order to efficiently simulate deformable containers, immersed in a fluid, we use a Lattice Boltzmann method as fluid solver, an explicit immersed boundary method for the coupling of the fluid and the membrane, and a finite element method for the computations of the membrane response to deformations. The surface of the particles is triangulated to allow efficient calculations of the deformations. The number of faces is in the range of 720 to 1280, which is sufficient to capture the studied deformations. For an overview of our method and membrane model, its relation to microscopic structure and the numerical evolution of the deformation gradient and its corresponding membrane forces we refer to \cite{Kruger2011,Kruger2013,Krugerthesis}. We will present our results in conventional lattice units and at the end of this section we will shortly outline the conversion to SI units.

\begin{figure*}
\includegraphics[scale=0.8]{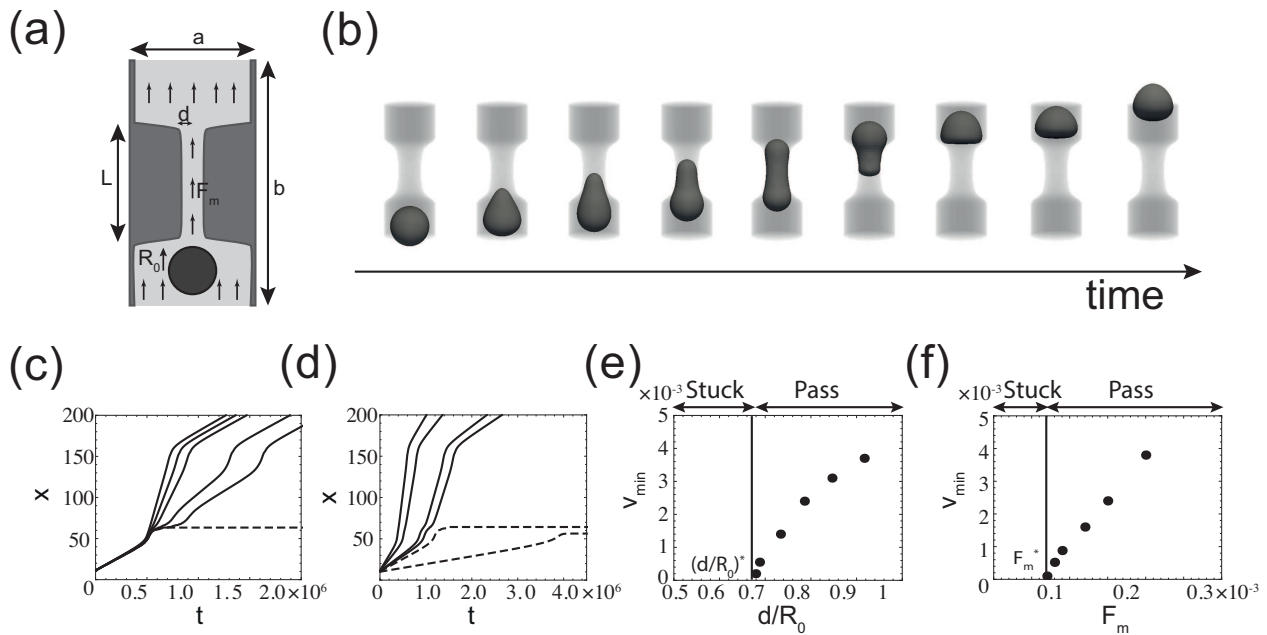}
\caption{ (a) Model system for the transport of containers through a narrow constriction of length $L$ and radius $d$. (b) A typical time sequence of the transport of a deformable container through a relatively short and narrow constriction. Position of the container's center of mass $x$ as function of time $t$ for (c) varying the relative radius of the constriction $d/R_0$ (solid line, { \it Pass}, $d/R_0 = 0.92$ - 0.68 and dashed line: {\it Stuck}, $d/R_0 = 0.67$) and a fixed applied body force $F_d=1.5 \times 10^{-4}$ and (d) where we vary the applied force $F_m$ (dashed line: {\it Stuck}, $F_m = 2 \times 10^{-5}$ and $6.5 \times 10^{-5}$ and solid line: {\it Pass}, $F_d =7 \times 10^{-5} - 2 \times 10^{-4}$) and fix $d/R_0 = 0.7$. In (e) we fix the applied body force $F_d = 1.5 \times 10^{-4}$  and in (f) we fix $d/R_0= 0.7$  and we calculate the minimal velocity of the particle during transportation. For the containers that remained stuck $v_{\rm{min}} =0$. In (c), (d), (e) and (f), the length of the constriction is $L= 100$ and the system size is $b=208$ and $a=52$ lattice units (See (a)). \label{fig:lbmodel} }
\end{figure*} 
 
Deviations from the equilibrium shape of the container incur an increase in the total energy, which we divide into three distinct contributions: i) energy due to in-plane strain: local contributions due to resistance to shear and to lateral dilatation, ii) energy due to out-of-plane bending, iii) energy due to global volumetric expansion or compression. The in-plane strain energy of an isotropic and homogeneous section of membrane is computed as
\begin{equation}
E_S = \int \epsilon_s d A,
\end{equation}
where $\epsilon_s$ is the surface strain energy density which depends on the principal stretches: the eigenvalues $(\lambda_1,\lambda_2)$ of the displacement gradient tensor $D$. In general, the strain energy density is a function only of the invariants $I_1 = \lambda^2_1 + \lambda^2_2-2$ and $I_2 = \lambda^2_1\lambda^2_2-1$, and any constitutive model is represented by a specific functional form for $\epsilon_s(I_1,I_2)$. As the deformations in biological cells are in general large, a linear stress-strain approximation is generally not justified. In our modeling, we implement therefore the nonlinear strain energy density proposed by Skalak \cite{Skalak1973} for biological membranes, valid for both small and large strains: $\epsilon_s= \kappa_s \left( I_1^2 + 2 I_1 - 2 I_2\right)/12 + \kappa_{\alpha} I_2^2/12$. $\kappa_s$ is the surface elastic shear modulus, and $\kappa_{\alpha}$ the area dilation modulus. For pure lipid bilayers, the in-plane behavior is liquidlike and $\kappa_s$ should be set to zero (in favor, technically, of a 2D viscosity multiplying the in-plane strain rate - we will, however, consider slow deformations and neglect viscous effects). For polymer capsules and even more complex mixtures of lipids, however, there will be contributions from the in-plane shear. Note that we neglect thermal area fluctuations - their effects are discussed in \cite{gompper1995}.

To account for the membrane bending, we recall the Helfrich bending energy \cite{helfrich1973},
\begin{equation}
E_B = \frac{\kappa_B}{2} \int(H-H_0)^2 d A ,
\label{eq:Fbend}
\end{equation}
where $\kappa_B$ is (the out-of-plane) bending modulus of the membrane, $H$ is the mean curvature and $H_0$ is the spontaneous curvature. The most general formulation of the Helfrich bending energy includes the Gaussian curvature term $\kappa_G \int K d A $, but this does not contribute to the overall energy provided no topological changes occur.

Finally, as the membrane is permeable to water, but not to ions, we associate an osmotic penalty for a deviation in volume given by
\begin{equation}
E_V = \frac{\kappa_V}{2} \frac{\left(V-V_0\right)^2}{V_0},
\end{equation}
where $V-V_0$ is the deviation in total volume and $\kappa_V$ is the volume modulus. 

We now use this model to address the question to what extent the force and the relative radius and length of the constriction affect the translocation of a deformable container (See Fig \ref{fig:lbmodel} (a)). In particular, we focus on a container with given mechanical properties (fixed $\kappa_V = 1$, $\kappa_{\alpha}=0.018$, $\kappa_s=0.5$ and $\kappa_b=0.05$), where we have used a dimensionless lattice constant, time-step and mass and set them all to unity, as well as the relaxation time. For the parameter values we choose here, the container is highly stretchable and strongly resists deviations in total volume. The bending contribution is relatively weak and as it is resistant to shear, it resembles a polymeric capsule rather than a bilayer membrane. As we show in the next section, however, the bending dominated limit and the stretching dominated limit show very similar behavior. In our simulations we assume that the fluid both inside and outside the capsule are Newtonian and both have the same properties, i.e. the same viscosity and density. In Fig. \ref{fig:lbmodel} (b) we show a typical time sequence of the transport of a deformable container through a constriction. 

To isolate the influence of neck size, relative to the radius of the container $d/R_0$, we consider a system with a neck length that is considerably longer than the size of the container within the constriction: $L= 100$ and measure the time-evolution of the position of the center of mass. We vary the radius of the container in the range $R_0 =$ 5.5 - 8.0 and fix the radius of the neck at $d = 5.5$ (See Fig. \ref{fig:lbmodel} (c)). The container is released at a distance 25 lattice units in front of the constriction and is propelled by a fixed body force $F_m = 1.5 \times 10^{-4}$ on all the fluid nodes. Similarly we also fix the size of the container and the size of the constriction $d/R_0=0.7$, and vary the applied body force $F_m=2 \times 10^{-5} -2 \times 10^{-4}$ and observe highly similar behavior (See Fig. \ref{fig:lbmodel} (d)). Below a threshold force $F_m^*$ and above a critical ratio $\left(d/R_0\right)^*$, the container remains {\it Stuck} in front of the constriction (dashed lines), and above $F_m^*$ and below $\left(d/R_0\right)^*$, the velocity within the constriction increases for increasing $F_m$, as can be seen in Fig.~\ref{fig:lbmodel} (c) and (d). 

To quantify the speed of the translocation, we extract the minimal velocity of the container in the constriction $v_{\rm{min}}$ as function of the relative size of the container $d/R_0$, as shown in Fig.~\ref{fig:lbmodel} (e), where a sharp transition between {\it Stuck} and {\it Pass} is found at a critical ratio $\left(d/R_0\right)^*$.  Above this value, $v_{\rm{min}}$ increases for increasing $d/R_0$. The exact value of $\left(d/R_0\right)^*$ depends on the magnitude of the applied body force as can be seen from the force dependence, where likewise we find that increasing the applied body force increases $v_{\rm{min}}$ (see Fig.~\ref{fig:lbmodel} (f)), and that below a threshold force $F_m^*$, the container remains {\it Stuck}. 

\begin{figure}
\includegraphics[scale=0.7]{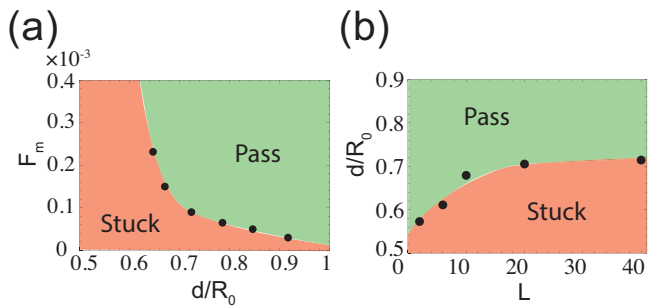}
\caption{ (a) Phase diagram indicating whether the container passes the constriction as function of the relative size of the constriction $d/R_0$ and the applied body force $F_m$. (b) Phase diagram as function of the neck length $L$ and the relative size of the constriction $d/R_0$, where $F_m = 1.5 \times 10^{-4}$. \label{fig:LBdia}}
\end{figure}

We combine the force and size dependence of the translocation into a single phase diagram, showing for which parameter values the container gets through the constriction ({\it Pass}) or remains stuck ({\it Stuck}) in Fig.~\ref{fig:LBdia} (a). The regime for $d/R_0< 0.6$ and $F_m>0.0002$, which is expected to show a power law-like behaviour of $F_m^*$ with decreasing $d/R_0$, at least for fluid vesicles \cite{gompper1995}, is presently inaccessible due to limitations in the simulation methods: velocities on the lattice nodes become too high and the weak incompressibility constraint may be numerically violated. 

As mentioned in the Introduction, both the radius and the length of the constriction are expected to affect the dynamics of the passage process. As we show in Fig.~\ref{fig:LBdia} (b), decreasing the length of the neck can considerably decrease the minimal relative radius of the neck $d/R_0$ through which the container can be forced. This is due to the fact that for shorter necks, one end of the container may already be exiting the constriction while the other end has not yet entered, allowing parts of the passage process to occur at considerably lower curvatures and thus to proceed more effectively. We discuss this in detail in Sec. \ref{sec3}.

All the results we have presented in this section were given in lattice units. These units can be converted to regular SI units. Although the scope of this section is to come up with a generic system and not to solve this problem for one particular system it is insightful to convert our lattice units to SI units for one particular case where the kinematic viscosity of the liquid equal that of water and the sound of speed is set to 27.8 m/s. For this particular case the lattice constant equals $1.25 \times 10^{-7} $m, which corresponds to a container of radius of the order of one micron. The resultant time step is then $ 2.6 \times 10^{-9} $s. As a result of this, the force densities $F_m$ on the lattice sites we have considered are in the order of $10^{9}\ \rm{N m^{-3}}$ or $10^{-9}\ \rm{N \mu m^{-3}}$, the area dilation modulus is $\kappa_{\alpha} = 5 \times 10^{-3} \rm{\ J m^{-2}}$ and the bending modulus is $\kappa_{B} = 2 \times 10^{-16} \ \rm{J}$. These values can be manipulated by varying the viscosity and the speed of sound of the medium; this has been discussed in more detail in Narv{\'a}ez et al. \cite{Narvaez2010}.

To conclude this section we mention that this particular simulation method poses some limitations as it does not permit a large range of mechanical properties of the deformable container to be studied and therefore, we are unable to simulate the bilayer limit, where the shear modulus is negligible and the area stretching modulus is very large. To access these regimes, we now present a tractable model for the two limiting cases of the capsule and the vesicle. 

\section{Limiting behaviors}\label{sec3}

\begin{figure*}
\includegraphics[scale=0.5]{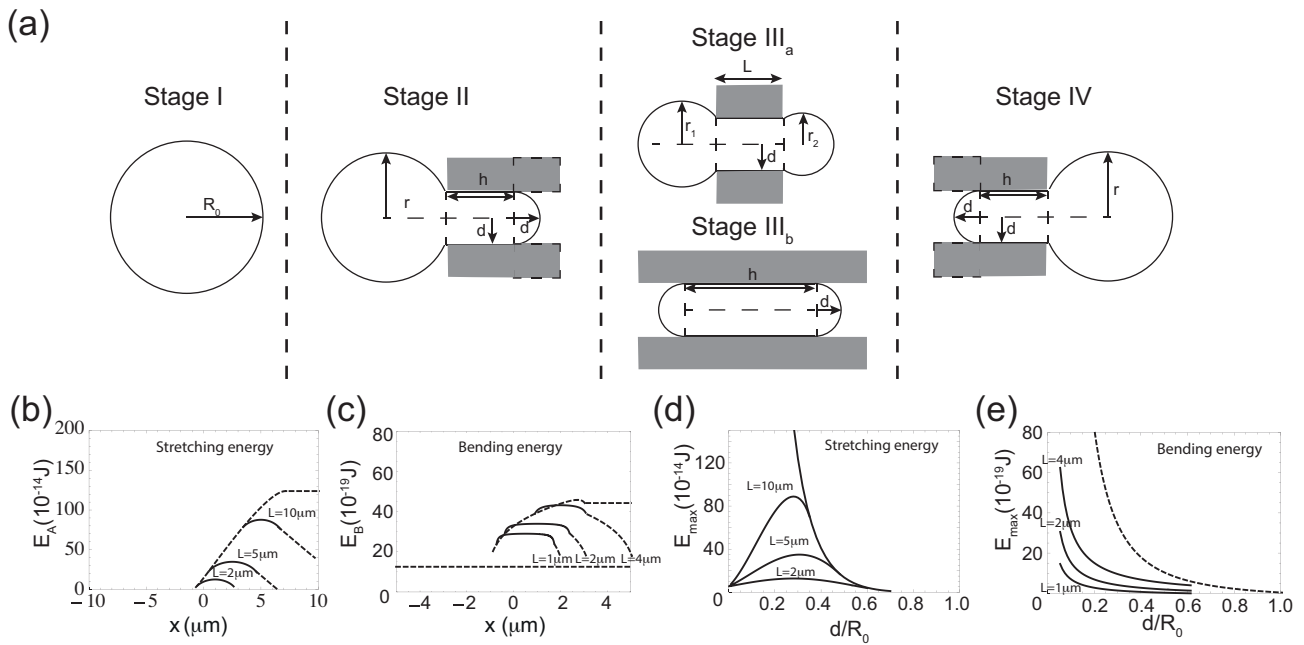}

\caption{ (a) Translocation sequence for the passage of a container through a narrow constriction. Stage (I): free capsule in solution (the reference configuration), stage (II): partial entry of the capsule into the constriction, stage (III$_a$): intermediate stage for short channels, or large containers, stage (III$_b$): intermediate stage for long channels, or small containers. Stage (IV): partial exit out of the constriction.  Stretching and Bending energy $E_A$ (d) and $E_B$ (e), respectively, as function of $x$ for various lengths of the constriction. Decreasing the length of the neck strongly diminishes the height of the energy barrier the container has to overcome. For (b) and (c) we assumed $R_0 = 1 \ \rm{\mu m}$, for (b) $d=0.2 \ \rm{\mu m }$ and $\kappa_A = 1 \ \rm{J m^{-2} }$ and for (c) $\kappa_B = 2 \times 10^{-19}  \ J$ and $d = 0.3 \ \rm{\mu m}$.  In (d) and (e) we show the height of the energy barrier as function of the radius of the constriction $d$ relative to the initial radius of the vesicle $R_0$, where for (d)  $\kappa_A = 1 \ \rm{J m^{-2} }$  and for (e) $\kappa_B = 2 \times 10^{-19} \ \rm{J}$. If we decrease the length of the constriction we find that the height of the barrier strongly diminishes. For an infinite neck length, the height of the barrier strongly increases for decreasing $d/R_0$. For finite neck lengths, however, there is a decrease at low $d/R_0$ in the stretching energy (d).\label{fig:elmodel}}
\end{figure*} 

In this section we consider two limiting cases of the translocation of a deformable container, the stretch-dominated and the bend-dominated. The geometry we consider is shown in Fig.~\ref{fig:elmodel} (a). We presume the dynamics to be determined by a balance of forces between a coarse grained hydrodynamic drag $F_d\left(  \dot{x}(t) \right)= 6 \pi \eta R_0 \dot{x}(t)$, with $\eta$ the dynamic viscosity of the fluid, $R_0$ the equilibrium radius of the container, and $\dot{x}(t)$ the instantaneous velocity of the containers center of mass, a driving force, associated for instance with the pulling by molecular motors, $F_m$, and a force opposing the motion due to the increase in membrane energy $F_g$. The latter, in the two limits, can be determined by calculating the derivative with respect to the center of mass position of either the global stretching energy $E_A$ for the stretch-dominated limit or the bending of the surface $E_B$ for the bend-dominated limit, such that
\begin{equation}
F_g \left(x(t)\right) - F_d\left( \dot{x}(t) \right) + F_m=m \ddot{x}(t).
\label{eq:motion}
\end{equation}
where $m$ is the mass of the container and $x(t)$ is the position of mass of the center of mass which for the non-spherical particles is calculated by summing the ``weighted" contributions of the subunits and assuming a homogeneous density. In the remainder of this section we will assume the mass to be $m = 4 \pi \times 10^{-18} $ kg and the dynamic viscosity of the fluid $\eta = 10^{-3} \ \rm{N s/ m^2}$. In the limit of low Reynolds number, Eq. \ref{eq:motion} reduces to  $F_g \left(x(t)\right) + F_m = 0$. We should note that the force we apply in this theoretical model is applied to the container and not to the fluid as we did in the simulations. Therefore, the numerical value of the force is actually significantly lower compared to that in the simulations.

We consider a spherical deformable container with radius $R_0$ that is transported through a cylindrical constriction with radius $d$ and length $L$ as depicted in Fig.~\ref{fig:elmodel} (a) and distinguish several distinct stages of the process: stage (I) is the free capsule in solution - the reference configuration for the container shape. During this stage the deformation force, $F_g$, acting on the container is zero and the motion is determined by a competition between driving and drag force. Stage (II): partial entry of the capsule into the constriction. Here, the hydrodynamic drag force is much smaller than the deformation force acting on the container.  Stage (III$_a$): intermediate stage for short channels, or large containers. Stage (III$_b$): intermediate stage for long channels, or small containers and finally stage (IV): partial exit out of the constriction. Stage (III$_b$) is only encountered when the volume of the container is larger than the total volume inside the constriction. We will now consider the stretch- and bending-dominated regime of this translocation. 

\subsection{Stretch-dominated (capsule) limit}\label{capsulelimit}

In the stretch-dominated regime, we assume that the total volume of the container is conserved ($V=V_0,\ E_V=0$) and that, given a certain global stretching modulus $\kappa_A$, the surface of the container is allowed to stretch. For simplicity we will account for a global energy penalty, associated with stretching:
\begin{equation}
E_A = \frac{\kappa_A}{2} \frac{\left(A-A_0\right)^2}{A_0} ,
\label{eq:Fstre}
\end{equation}
where $A-A_0$ is the deviation in total surface and $\kappa_A$ a global stretching modulus. In the limit of small and uniform stretch ($\lambda_1 = \lambda_2$) and zero shear modulus $\kappa_s$, $\kappa_A$ can be related to the more general local area dilatation modulus $\kappa_{\alpha}$: $4 \kappa_{\alpha}/3 = 2 \kappa_A$. We will refer to containers in this regime as {\it capsules}.  This approach enables us to calculate analytically the height of the elastic energy barrier due to stretching of a container. As shown in Eq. \ref{eq:Fstre} we need to calculate the difference in total surface area for these three situations, illustrated in Fig.~\ref{fig:elmodel} (a). In Appendix A we detail the calculations of the total stretching energy as function of the center of mass of this system. Fig.~\ref{fig:elmodel} (b) collects the results, showing the elastic energy $E_A$ as function of the position of the center of mass where we fix the radius of the constriction and the stretching modulus and vary the constriction length. We find that upon decreasing the length of the constriction, the height of the energy barrier decreases considerably as can be seen in Fig.~\ref{fig:elmodel} (b) and (d). The height of the barrier is determined by the most stretched configuration that is encountered during the passage.  For large containers (or smaller channels) the most stretched state is attained at the moment during stage (III$_a$) when $R_1=R_2$. If the length of the constriction is greater than that of the capsule, the spherocylindric capsule (stage (III$_b$) in Fig.~\ref{fig:elmodel} (a)) is the state with maximal surface area and thus the maximal elastic energy. For shorter neck lengths, there is a single maximum set by the symmetric intermediate stage (III$_a$) shape.

The height of the energy barrier is thus proportional to the square of the deviation in total surface between the stage (III$_a$) shape and the sphere. If we calculate the height of this barrier as function of the relative size of the neck $d/R_0$ we find, for the infinitely long constriction, that upon decreasing $d/R_0$ the height of the barrier for $ d/R_0<0.2$, increases as $E_{\rm{max}} \sim \kappa_A \left( \left( R_0^3-R_1^3\right)/dR_0\right)^{2}$. If we now decrease the length of the constriction we find that for $d/R_0 \rightarrow 0$ the constricted capsule consists of two spheres with a total surface area equal to $4 \pi R_0^2/2^{2/3}$, which corresponds with two equally sized spheres. This limiting case has a smaller surface area than a system with a slightly larger $d/R_0$ and therefore a lower stretching energy. Therefore, there is a length of the constriction for which the stretching energy is maximal, and at which upon increasing and decreasing $d/R_0$, the height of the energy barrier decreases. In Fig.~\ref{fig:elmodel} (d) we plot the height of the barrier as function of $d/R_0$ for various neck lengths. Obviously, in realistic biological systems, this description would fail as there would be a very high bending involved with such thin necks.

\begin{figure*}
\includegraphics[scale=0.72]{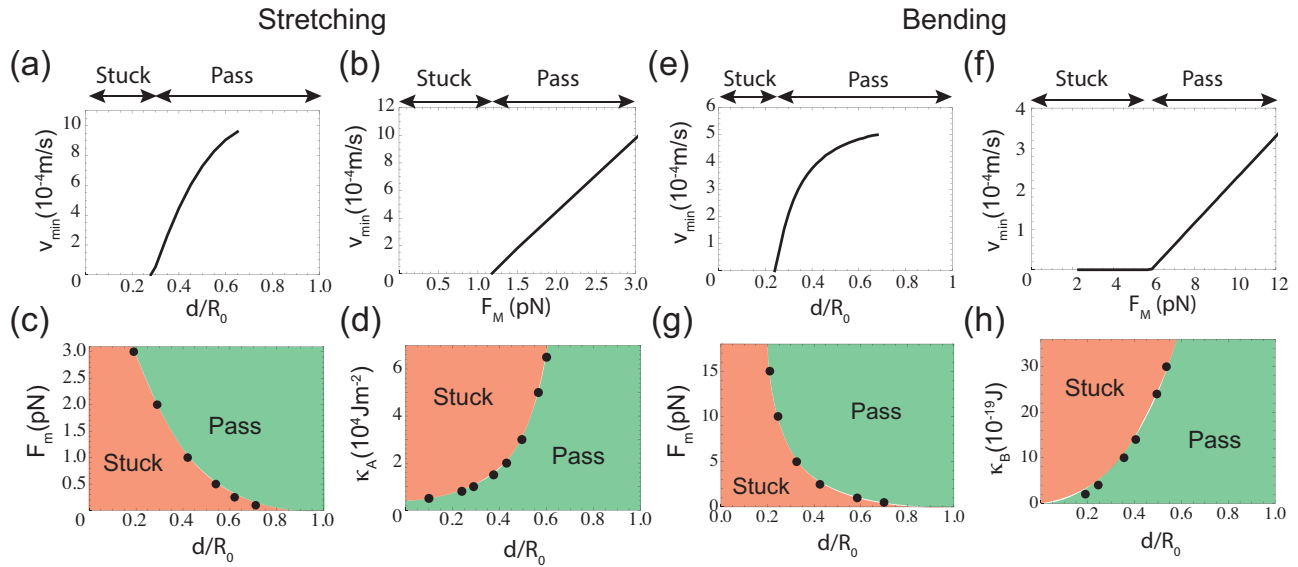}
\caption{ The minimal velocity during the entrence of the constriction $v_{\rm{min}}$ as function of (a) the relative radius of the constriction $d/R_0$ ($\kappa_A=10^5$ $\rm{J m^{-2} }$, $F_m= 2$ $\rm{pN}$) and (b) as function of the applied force $F_m$ ($\kappa_A=10^5 \rm{ J m^{-2}}$, $d/R_0= 0.4$). Phase diagram indicating if the stretch-dominated container {\it Passes} or gets {\it Stuck} inside the constriction as function of (c) $F_m$ and $d/R_0$ ($\kappa_A=10^5 \  \rm{J m^{-2}} $) and (d) $\kappa_A$ and $d/R_0$ ($F_m= 4$ pN). $v_{\rm{min}}$ as function of (e) $d/R_0$ ($\kappa_B=4 \times 10^{-19}  \ \rm{J}$ and  $F_m= 10$ pN) and (f) $F_m$ ($\kappa_B=8 \times 10^{-19}  \ \rm{J}$ and $d/R_0= 0.4$). Phase diagram indicating if a bend-dominated container  {\it Passes} or gets {\it Stuck} as function of (g) $F_m$ and $d/R_0$ ($\kappa_B=4 \times 10^{-19}  \ \rm{J}$) and (h) $\kappa_B$ and $d/R_0$ ($F_m= 10 $ pN).  These figures show a very generic phase behavior for both the stretch- and bend- dominated containers. \label{fig:phaseves}}
\end{figure*}

From the deformation energy as a function of position we determine $F_g={\dd E_A}/{\dd x}$, the force that opposes translocation due to the shape change. We solve the force balance (Eq. \ref{eq:motion}) to determine $x(t)$ and $\dot{x}(t)$ and extract the pass-stuck phase diagram. Similar to the simulations of Section II, we obtain $x(t)$ and $\dot{x}(t)$, from which we can calculate the phase diagram and minimal velocity $v_{\rm{min}}$ within the constriction. This point of minimal velocity corresponds to the point of maximal stretching force and we will use this quantity to characterize the motion of the container. We have performed the calculations of $v_{\rm{min}}$ for various radii of constriction and found that, for a given force, decreasing the size of the constriction decreases the minimal velocity inside the constriction. Eventually, at a critical radius $(d/R_0)^*$, this velocity will become zero and the container will get stuck in the constriction. Above this critical radius, the minimal velocity increases with increasing $d/R_0$ as shown in Fig.~\ref{fig:phaseves} (a). The occurrence of a critical threshold also holds for the driving force $F_m$, above which, the velocity increases linearly with the driving force $F_m$ (See Fig.~\ref{fig:phaseves} (b)). The dependence of the minimal velocity on driving force and $d/R_0 $ allow us to create a phase diagram indicating whether a capsule gets through the constriction or not: this is presented in Fig.~\ref{fig:phaseves} (c). This phase diagram indicates the critical force $F_m$ necessary to translocate a container of radius $R_0$ through a constriction with size $d$, for a given stretching modulus $\kappa_A$. If we now fix the driving force and vary the stretching modulus $\kappa_A$ and $d/R_0$, we obtain a similar phase diagram for the critical $\kappa_A$. Obviously, the minimal size of the constriction through which a container would pass decreases strongly with decreasing modulus (See Fig.~\ref{fig:phaseves} (d)). While for biological membranes the elastic parameters are largely fixed, in synthetic systems one may have some control over the area elastic properties.

\subsection{Bend-dominated (vesicle) limit}

We now analyze the opposite limit, where the container has very limited opportunity to stretch, and the elastic energy is dominated by the bending contribution $E_B$. This would resemble more closely a biological membrane, whose internal volume may adapt due to the relatively high permeability to water of lipid bilayers. Though there may be some areal extension, we will assume its energy is negligible compared to the bending contributions. As our reference configuration, we take again the spherical vesicle, and the various stages of translocation are the same as in Fig.~\ref{fig:elmodel} (a). The calculation of the elastic energy for this type of container is highly similar to that of the stretchable container -  we refer to Appendix B for the details and summarize only our main findings here.

Fig.~\ref{fig:elmodel} (e) shows the elastic energy for a bend-dominated container for a constriction with finite and infinite length. The transition from stage (I), the free container, to the stage (II) is no longer continuous. This jump in the bending energy is an artifact of our simplified setup, as our model does not resolve the continuous transition from situation (I) to (II). Once the container enters the constriction, its energy increases until it reaches a maximum. For larger vesicles, or short channels, this point corresponds to the symmetric configuration during stage (III$_a$) when $R_1=R_2$, provided the length of the constriction is short enough that the container can span both ends of the constriction. For smaller vesicles, or longer channels, the stage (II) bending energy continues to increase until the vesicle has completely entered the constriction to reach stage (III$_b$) - a spherocylinder completely inside the channel. Past this point, both stage (III$_a$) and (III$_b$) develop into stage (IV), where the energy decreases in the inverse manner that it rose in stage (II). 

If we now compute the maximal height of the energy barrier $E_{\rm{max}}$ in Fig.~\ref{fig:elmodel} (g) as function of the radius of the constriction, we find that it increases strongly for decreasing radius and length of the constriction. For long channels, we find a scaling regime where $E_B \sim \kappa_B (d/R_0)^{-2}$ which is highly similar to what we found for the stretchable container in Fig.~\ref{fig:elmodel} (c). There is, however, one notable difference: the barrier height does not display the maximum we find in the capsule limit. This may be understood from the fact that the bending energy diverges for small $d$, whereas $E_A$ does not.

Next, we use the equation of motion (Eq. \ref{eq:motion}) to obtain the dependence of the minimal velocity during the passage through the constriction as a function of the relative radius of the constriction $d/R_0$ (Fig.~\ref{fig:phaseves} (e)) and the force applied to the vesicle $F_m$ (Fig.~\ref{fig:phaseves} (f)). Figs.~\ref{fig:phaseves} (e) and (f) reveal similar behavior as for the stretch-dominated container: below a threshold force, the container gets stuck and above this critical force, its velocity increases linearly with increasing force $F_m$. If the size of the neck, relative to the size of the vesicle, is decreased below a critical ratio $d/R_0$, the vesicle gets stuck. Above this value the minimal velocity increases as shown in Fig. \ref{fig:phaseves} (e). 

We summarize in Fig.~\ref{fig:phaseves} (g) and (h) the force-dependence of the translocation in a phase diagram, indicating under which combinations of parameters the vesicle gets through the constriction and when it does not. This phase diagram indicates the critical force $F_m$ necessary to transport a vesicle of radius $R_0$ through a constriction with radius $d$. If we now fix the driving force and vary the bending modulus $\kappa_B$ and $d/R_0$, again at fixed area increasingly small bending moduli are required to pass through the channel. Overall, the results are very similar to those in the stretch-dominated limit and those observed in the simulations presented in Section II. 

\section{Conclusions}

In this paper, we have sought to address the question of how passage dimensions, container mechanics and external forcing together determine whether or not a container will pass through a narrow constriction, and if it does - how fast it does so. We have shown that by varying the size of the container relative to that of the neck and by regulating the force that is exerted on the container, the system may be biologically or physically controlled to, for instance, switch between a state where the container remains {\it Stuck} in front of the neck and a state where the container passes through the neck. Both these states possess some biological significance. 

We have presented the results of Lattice-Boltzmann simulations, supported by two limiting simplified theoretical models. Although a quantitative comparison between the simulations and the theoretical is difficult to establish as a result of, among others, the dependence on the exact driving mechanism we find that even while the energetics of highly stretchable containers is very different from that of containers that are bend-energy-dominated, the resulting phase diagram, in terms of {\it Stuck} vs. {\it Pass}, is very similar in both cases, suggesting some universality between both limits. We focus on the scaling and a qualitative analysis of this problem in the regime of low flow rates, justifying the fact that we neglect both the membrane viscosity as well as the solvent viscosity in our models. At high flow rates, the imposed strain rate the membrane experiences may lead to significant contributions from both the viscosity of the membrane as well as the viscosity of the solvent, and different behaviors from those we describe here are to be expected.

Nonetheless, our modeling allows us to address some of the questions we have raised in the introduction: whether typical cellular force levels are sufficient to effect translocation in typically dimensioned vesicles and constrictions, and whether it is feasible for a cell to switch between pass and stuck by controlling this force. In order to do so, we must quantify the position within the {\it Stuck}/{\it Pass} phase diagram for a typical biological cell. We may use our results to provide some quantitative insight into the passage of biological vesicles into thin necks, such as it occurs in the dendritic spines mentioned in the introduction. In Fig.~\ref{fig:phaseves} (g) we show the phase diagram as function of the minimal force that a molecular motor has to exert vs. the size of the neck, where we have substituted typical values of relevant parameters for a recycling endosome, which has an equilibrium radius of $1 \ \rm{\mu m}$ \cite{Park2006, Wang2008}, and the bending modulus of a typical vesicle $\kappa_B = 4\times 10^{-19} \ \rm{J}$ \cite{Semrau2008}. Our analysis shows, that the range of forces necessary to transport this container through a typical dendritic spine neck, which has a radius of between $0.2-0.6 \ \rm{\mu m}$, is on the order of a few to tens of pN. A typical myosin motor can exert forces of 5-6 pN \cite{Fisher1999, Kolomeisky2007}. The dimensions of the dendritic neck - a very typical channel motif in cells - thus require one to a few motors to translocate vesicle-bound cargo, confirming that motors are eminently capable of producing the requisite forces to selectively translocate or immobilize vesicles in the neck, and to switch between these modes. We will note that although the translocation is dominated by a competition between deformation energy and forcing, it cannot be expected from this simple model to accurately capture the exact forcing involved in motor transport - indeed, pulling by motors bound to cytoskeletal polymers arranged mostly close to the cell membrane in the neck is likely to affect the shape of the endosome. These additional contributions are, however, unlikely to dominate; the principal bending energy contribution still comes from the highly elongated transitional shape during stage III$_b$.

In future simulations we will include a more realistic driving mechanism to the container such that the Lattice-Boltzmann simulation may be used to capture the dynamics of the problem in even more detail. We expect our quantitative results to depend on the precise driving mechanism, which will be addressed in future research. In addition to this, we are investigating to what extent the actin network within the neck of the dendritic spines hinders the transport of membrane containers \cite{Hotulainen2010}.
\section*{Acknowledgments}
We thank Timm Kr{\"u}ger for valuable discussions. This work was supported by funds from the Netherlands Organization for Scientific Research (NWO-FOM) within the programme "Barriers in the Brain: the Molecular Physics of Learning and Memory" (No. FOM-E1012M) and the VIDI grant 10787 of Jens Harting "Dense suspensions in medicine and industry".

\appendix

\section{Stretch-dominated limit}
In this Appendix we outline the calculations of the energetics involved in the stretch-dominated regime. As mentioned in the main text we assume that for the container in the stretch-dominated limit that the total volume is conserved, and that the total surface determines the stretching energy as shown in Eq. \ref{eq:Fstre}. The surface of the container in the constriction can be divided in three parts (see Fig.~\ref{fig:elmodel} (a) for parameters), and in stage (III$_a$) the total surface area of the capsule is computed to be 
\begin{small}
\begin{equation}
\begin{split} 
 &A_{tot} = 4 \pi R_1^2 - \pi \left( d^2 + \left( R_1- \sqrt{ \left( R_1^2 - d^2\right) } \right) \right)^2 \\
 &+ 2 \pi  l d +  4 \pi R_2^2 - \pi \left( d^2 + \left( R_2- \sqrt{ \left( R_2^2 - d^2\right) } \right) \right)^2\ .
 \end{split}
\end{equation}
 \end{small}
The radius of the first spherical cap $R_1$ is related to that of the second spherical cap $R_2$ via total volume conservation:
\begin{small}
\begin{equation}
\begin{split} 
 V_{tot} &= \frac{4 \pi R_0^3}{3}  \\ 
 &=  \frac{4 \pi R_1^3}{3} - \frac{\pi h^2}{3} \left( 3 R_1 -R_1 - \sqrt{R_1^2 - d^2}\right) \\
 &+ \pi d^2 l +\frac{4 \pi R_2^3}{3} - \frac{\pi h^2}{3} \left( 3 R_2 -R_2 - \sqrt{R_2^2 - d^2}\right),
 \end{split}
\end{equation}
\end{small}
where $V_{tot}= 4 \pi R_0^3/3$ is conserved. To calculate the evolution of the stretching energy, we identify the position of the center of mass of this system, then calculate the shape of the system and the corresponding area deviation. In stage (I), where we have a spherical capsule at its equilibrium radius $R_0$, the total stretching energy is 0 as $A=A_0$. The deviation in total surface area and the corresponding stretching energy of the stages (II) and (III$_a$/III$_b$) are calculated assuming total volume conservation. In stage II, we divide the membrane shape into three domains: a partial sphere of radius $R$ outside the constriction, and a (truncated) spherocylinder with length $h$ and radius (both of the cylindrical section and the spherical cap) equal to the radius of the channel, $d$. In stage (III$_a$), likewise, we distinguish three domains: a spherical cap with radius $R_1$ outside the entry, a cylindrical tube with radius $d$ and length $h$ inside the channel, and a spherical cap with radius $R_2$ outside the exit. In stage (III$_b$) the shape is a spherocylinder with length $h$ and all radii equal to $d$.

\section{Bend-dominated limit}

In this appendix we outline the calculations of the energetics involved in the bend-dominated regime. For the initial state we consider a spherical vesicle that has an initial radius $R_0$ before it enters the constriction. Its bending energy is given by $ E_B=  2 \pi \kappa_B$, independent of the radius. In stage (II), we consider the entrance of the vesicle in the constriction. Again, the complete shape is divided into two, possibly distinct, spherical domains and the cylindrical part. Note that in contrast to the capsule, the total surface area of the vesicle is conserved: $A_{tot} = A_{I} + A_{II}+ A_{III}$, where: $ A_{I}\left( R_1, d \right) = 4 \pi R_1^2 - \pi (d^2 +\left(R_1^2+\sqrt{R_1^2-d^2} \right)$,  $A_{II}\left( h, d \right) = 2 \pi d h$ and $ A_{III}\left( R_2, d \right) = 4 \pi R_2^2 - \pi (d^2 +\left(R_2^2+\sqrt{R_2^2-d^2} \right)$. The volume of the vesicle is allowed to increase during the translocation.

The bending energy associated to the vesicle in stage (III$_a$) or (III$_b$) can be approximated by the sum of the three separate contributions:
 \begin{equation}
 E_B = \kappa_B/2 \left(\frac{A_{I} (R_1, d)}{R_1^2} + \frac{ A_{II} (h, d)}{4 d^2}  + \frac{A_{III}(R_2,d)}{R_2^2}\right),
\label{eq:free}
 \end{equation}
 where $A_{I}$ is the surface of the spherical cap with radius $R_1$, $A_{II}$ the cylindrical part with radius $d$ and length $h$ and $A_{III}$ the spherical cap with radius $R_2$. 
 
As mentioned in the main text, by assuming that the total surface area is conserved, on can relate $R_1$ and $R_2$ to the initial radius $R_0$ and the position of the center of mass $x$. As depicted in Fig.~\ref{fig:elmodel} (a) we consider two separate situations, (III$_a$) and (III$_b$). in the first, we assume a neck with short length $L < R_0$ and the situation where $L \ll R_0$. To calculate the bending energy of the first situation we fix $R_2=d$. By fixing the total area we obtain the following relation between the radius of the first spherical part $R_1$ and the length of the cylindrical domain $h$:
\begin{equation}
R_1(h) = \frac{A_{tot}- 2 d^2 \pi -2 \pi d h}{\sqrt{4 \pi A_{tot} - 12 \pi^2 d^2 - 8 \pi^2 h d}}.
\end{equation}
We use this condition to solve Eq. \ref{eq:free} as long as $h<L$. If $h>L$, we fix $h=L$ and by conserving the total area one can determine a relation between $R_1$ and $R_2$.
Assuming that the volume enclosed by the neck is $2 \pi d L \ll 4 \pi R_0^2$ one should consider an extra situation which is the vesicle completely inside the neck - stage (III$_b$). The calculation is a straightforward extension of the previous setting $R_1 = R_2 = d$. This yields the following relation between the length of the neck and the total area of the vesicle
\begin{equation}
h(d) = \frac{A_{tot}- 4 d \pi d^2}{2 \pi d}.
\end{equation}

\end{document}